*Deterrence and Prevention-based Model to Mitigate Information Security Insider Threats in Organisations*

Nader Sohrabi Safa[*a], Carsten Maple[b], Steve Furnell[c], Muhammad Ajmal Azad[d], Charith Perera[e]
Mohammad Dabbagh[f], *Mehdi Sookhak*[g]

School of Computing, Electronics and Mathematics, Coventry University, United Kingdom[a]
Cyber Security Centre, WMG, University of Warwick, United Kingdom[b]
School of Computing, Electronics and Mathematics, University of Plymouth, United Kingdom[c]
Department of Electronics, Computing, and Mathematics, University of Derby, United Kingdom[d]
School of Computer Science and Informatics, Cardiff University, United Kingdom[e]
Department of Computing and Information Systems, School of Science and Technology, Sunway University, Malaysia[f]
Polytechnic School, Arizona State University, USA[g]

*Abstract*

Previous studies show that information security breaches and privacy violations are important issues for organisations and people. It is acknowledged that decreasing the risk in this domain requires consideration of the technological aspects of information security alongside human aspects. Employees intentionally or unintentionally account for a significant portion of the threats to information assets in organisations. This research presents a novel conceptual framework to mitigate the risk of insiders using deterrence and prevention approaches. Deterrence factors discourage employees from engaging in information security misbehaviour in organisations, and situational crime prevention factors encourage them to prevent information security misconduct. Our findings show that perceived sanctions certainty and severity significantly influence individuals' attitudes and deter them from information security misconduct. In addition, the output revealed that increasing the effort, risk and reducing the reward (benefits of crime) influence the employees' attitudes towards prevent information security misbehaviour. However, removing excuses and reducing provocations do not significantly influence individuals' attitudes towards prevent information security misconduct. Finally, the output of the data analysis also showed that subjective norms, perceived behavioural control and attitude influence individuals' intentions, and, ultimately, their behaviour towards avoiding information security misbehaviour.

## *1.* Introduction

Several reports show that a significant portion of information security breaches originate from insiders [1-3]. Confidentiality of information, particularly when relating to industrial design, infrastructure control, experts' information, organisational information assets and so forth, is an important matter. In addition, information is a competitive resource in many organisations, and information leakage has serious consequences for firms, such as reputational damage, loss of revenue, loss of intellectual property, a reduction in productivity and competitive advantage, costs arising, and, in the worst-case scenario, bankruptcy [4, 5]. Information leakage refers to the accidental or deliberate transfer of information to an unauthorised person or persons within or outside an organisational boundary [6, 7]. It is acknowledged that technology alone cannot ensure a secure environment for information assets; the human aspects of information security should also be taken into consideration [8-10]. The confidentiality of information and data has managerial aspects. Different experts have presented various approaches to protect information with regard to human aspects. Siponen, Adam Mahmood [11], Ma, Jiang [12] and Sohrabi Safa, Von Solms [13] consider the idea that complying with organisational information security policies and procedures (OISPs) is an effective and efficient avenue for mitigating information security breaches. Information security knowledge sharing has been presented as another approach that decreases information security threats whilst increasing the knowledge and awareness of employees in the organisation [14, 15]. Conscious care behaviour, which is based on information security awareness and experience, has been presented as another effective approach that mitigates human mistakes in the domain of information security [16]. However, this research aims to investigate the effect of deterrent and preventative factors on employees' behaviour in order to decrease insider threats in organisation.

Crime is reduced when no motivation exists [17]. In many studies, motivation for crime has been mentioned as being an important factor [18, 19]. This is the salient factor that we suggest is used to reduce information security misbehaviour in organisations. Motivation can explain individuals' behaviour in many cases. Motivation is what encourages an individual to behave in a specific way or incline towards a certain kind of behaviour. Motivation creates a direction for a behaviour [20]. Wang and Hou [21] investigated the effect of altruism, and hard and soft rewards as motivational factors that encourage knowledge sharing among employees. In this research, the term 'hard rewards' refers to benefits such as financial rewards in an organisation, while 'soft rewards' relate to the emotional pleasure such as relationships with significant others or personal reputation. Shibchurn and Yan [20] explored the effect of extrinsic and intrinsic motivations on the exposure of information on social networks. The results of their study revealed that there are positive correlations between intrinsic and extrinsic motivations with information disclosure intention. Several studies investigated the effect of social bond factors – attachment to organisation, commitment to organisational aims, involvement in particular activity such as information security and personal norms – as motivational factors that encourage employees to comply with OISP [13, 22].

It is acknowledged that sanctions, as well as rules and regulations, constitute formal controls. Formal controls are intended to influence individuals' behaviour in such a way as to prevent deviant behaviour [23, 24]. The General Deterrence Theory (GDT) explains how people avoid deviant behaviour in the context of a society. GDT is based on negative motivations innate in formal sanctions. This theory encompasses two important elements – sanction certainty and sanction severity. 'Sanction certainty' refers to the belief that individuals' misbehaviour will be detected. 'Sanction severity' refers to the fact that the deviant behaviour leads to harsh punishment [25]. The punishment mechanism encompasses jailtime, fines, dismissal or denunciation. Both sanction certainty and severity negatively influence the intention of individuals to engage in misbehaviour in organisations. GDT is amongst the most favoured theories in the information security realm [26, 27]. The motivational and deterrence aspects of GDT are two important parts of the research model developed in this study.

In this paper, the theoretical background with a description of the Deterrence Theory (DT), Situational Crime Prevention Theory (SCPT) and the Theory of Planned Behaviour (TPB) are explained in section two. The research conceptual model and its hypotheses are described in section three. The methodology of the research, data gathering and demography of the participants are presented in section four. The results of the statistical analysis, measurement model (MM) and structural model (SM) are discussed in section five. The contribution and implementation of the research are illustrated in section six. Finally, the conclusion, limitations and topics of future work are explained in section seven.

## *2. Theoretical background*

This study aims to decrease insider threats using a novel approach – deterrence and opportunity reduction for information security misbehaviour. We synthesised the DT and SCPT in order to examine how to change the attitude and mindset of employees with a view to preventing misconduct in the domain of information security in organisations. In addition, the TPB explains how affective factors influence employees' behaviour. We believe that this theoretical background together with a comprehensive literature review increase the reliability of the research model. Both the Deterrence and Situational Crime Prevention factors are aligned with each other and have the same effect on individuals' attitude. These two theories, alongside the TPB, show the complete chain of behaviour change, and explain how we can improve employees' information security behaviour and mitigate the risk of information security breaches.

**2.1. General Deterrence Theory**

The General Deterrence Theory (GDT) describes human behaviour and decisions in terms of minimising their cost and maximizing their benefit to the individual. Losing reputation, competitive advantage, productivity and profit can be consequences of employees' who, through their behaviour, threaten the availability, confidentiality and integrity of the information assets in organisations. It is acknowledged that deterrent approaches, such as disincentives and sanctions influence the direction of individuals' behaviour towards avoiding certain actions in a community. The effectiveness of such disincentives is based on the certainty and severity of sanctions [28]. If an offender realises that his or her criminal act will be detected (sanction certainty) and that the authority will consider harsh punishment, such as a fine, jailtime, dismissal, denunciation, or some other forms of punishment (sanction severity), he or she will not engage in deviant behaviour [29]. The GDT has been applied as an effective and efficient approach to comply with OISP [30]. In this research, the GDT has been used to show how sanction certainty and severity influence the attitude and intention of employees with the effect of preventing deviant behaviour in the domain of information security.

## 2.2. Situational Crime Prevention Theory

Motivation and opportunity are two important factors in the formation of different crimes. The Situational Crime Prevention Theory (SCPT) explains how we can decrease motivation and opportunity in order to reduce criminal activities or delinquent behaviour [31]. The SCPT is a common approach that mitigates motivation and opportunity for many types of crime. In this regard, opportunity reduction mechanisms have been acknowledged as being an effective and efficient approach toward reducing delinquent behaviour in many communities [32]. The SCPT helps management to design an environment to control delinquent behaviour or crime based on different perspectives. This approach can be applied in various environments and contexts, such as organisations, schools, social networks, ecommerce and other similar communities. Available opportunities and rationalisation encourage offenders to conduct illegal activities or crimes. It is acknowledged that if offending is difficult, the motivation to perpetrate delinquent behaviour or crime will reduce. The benefit and cost of the offender's behaviour are important to them; hence, the benefit and cost of their actions influence their decision to engage in delinquent behaviour [33]. The SCPT mitigates delinquent behaviour by making crimes more difficult and risky, and reduces the rewards which constitute the output of the crime, as well as reducing the excuses and provocations to prevent rationalization for perpetrating crimes. The situational crime prevention mechanism has been applied as an effective and efficient approach to mitigate insider threats in organisations in this research.

## 2.3. Theory of Planned Behaviour

Individuals' behaviour is influenced by their beliefs. Ajzen and Madden [34] proposed the Theory of Reasoned Action (TRA) to explain human behaviour based on intention, subjective norms and attitudes. The TRA was further developed by adding perceived behavioural control to better explain individuals' behaviour. The TPB encompasses attitude, perceived behavioural control, subjective norms, and intention. According to the TPB, if people evaluate a behaviour positively (attitude), and if they think that other important persons want to conduct their behaviour in the same way (subjective norm), and if they have the ability and potential to perform it (perceived behavioural control), then they have a stronger intention to conduct the behaviour. Clearly, the TPB can explain human behaviour in various fields, such as public relationships, organisational behaviour, advertising, healthcare, and campaigns. Cox [4] explained information security awareness and assurance using TPB. In another study, Ifinedo [35] explained compliance with OISP by applying TPB. In the present research, TPB has been used in order to develop a conception of how to mitigate the risk of information security misconduct in organisations. Figure 1 shows the research model and theories in a concise form.

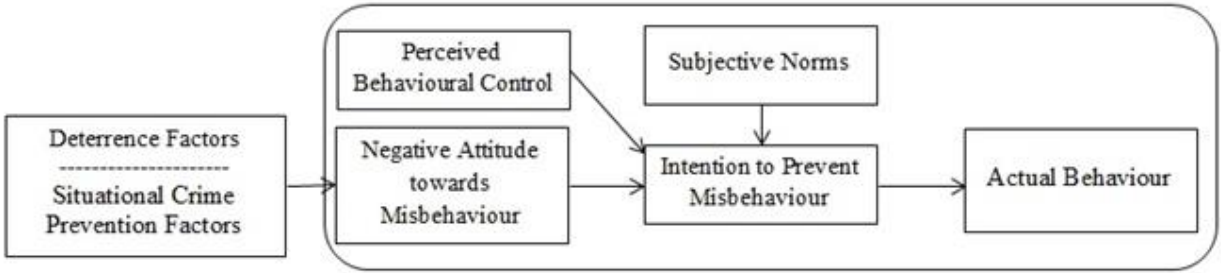

Figure 1: Research model

## 3. Research model and hypotheses

This study aims to investigate the effect of sanction certainty and severity as deterrent factors on the one hand, and the effect of increasing the effort and risk, reducing the rewards and provocations, and removing excuses as situational crime prevention factors, on the other, on employees' attitude towards preventing misbehaviour in the domain of information security. The Deterrence and Situational Crime Prevention Theories alongside the Theory of Planned Behaviour depict the effect of these factors on employees' attitudes, intention, and, ultimately, behaviour. Perceived behavioural control and subjective norms also influence the intention of employees towards changing their behaviour, based on the TPB. We can see a complete chain of behaviour formation in the research model.

### 3.1. Perceived sanction certainty and severity

It is acknowledged that deterrence factors negatively influence the decision of individuals to be involved in crime. The GDT has been frequently applied to explain human behaviour in various disciplines. The certainty and severity of punishment influences the minds of individuals and their decision to commit crime or engage in delinquent behaviour. Based on Deterrence Theory, to some degree, human behaviour is rational and can be influenced by negative incentives inherent in formal sanctions [36]. 'Sanction certainty' refers to the belief of an individual that his or her delinquent behaviour will be detected by the relevant authority, while sanction severity relates to the belief that he or she will be punished because of his or her delinquent behaviour [37]. Henle and Blanchard [38] showed that sanction certainty and severity decrease cyber loafing and abuse of organisational equipment. Siponen and Vance [39] presented the effect of these two factors on employees' compliance with the OISP. This research strives to investigate the effect of sanction certainty and severity on employees' misbehaviour in the domain of information security. Hence, the following hypotheses are presented:

*H 1:* Sanction certainty positively influences employees' attitudes towards preventing delinquent behaviour in the domain of information security.

*H 2:* Sanction severity positively influences employees' attitudes towards preventing delinquent behaviour in the domain of information security.

## 3.2. Increase the Effort

The difficulty in carrying out an action influences an individual's attitude and decision to pursue their plan. This approach can be applied in order to increase the difficulty of executing violations by employees in organisational environments [40]. Account policies and closing the doors of unauthorised data exfiltration, the monitoring of facilities, and the strong enforcement of password and access controls are examples of organisational actions that make information security violation difficult for offenders [33, 41]. A combination of different methods can be more effective in this regard. Authentication should be supplemented by access control to be more effective in controlling access to the system or data in organisations. However, traditional access controls, such as role-based access controls, are vulnerable to insider threats, unless the access control is updated frequently. Hence, kinds of access control in addition to Finger-grained authentication may be an effective strategy to increase the effort expended for information security misbehaviour [42]. Therefore, we postulate that:

*H 3:* Increasing the effort for information security misconduct positively influences employees' attitude towards preventing delinquent behaviour in the domain of information security.

## 3.3. Increase the Risk

Risk is the potential of gaining or losing valuable things. Losing social status, financial wealth, health and reputation are examples of risks that are outcomes of the behaviour of individuals. People think about the consequences of their actions before conducting them; the measure of risk influences their attitude towards and decision concerning the engagement in a violation or crime [29]. In other words, increasing the risk is associated with the increased probability of identifying the offender, detection of the violation by the authority, or apprehension resulting from malfeasance [19]. An event management system, auditing and monitoring the actions of individuals, using a log correlation engine, reducing anonymity, and monitoring and controlling remote access can increase the risk for employees who engage in information security misbehaviour. Insider reporting is another effective approach that increases the risk for offenders and improves information security surveillance. The prediction of future incidents by investigating similar previous incidents also increases the risk for offenders and decreases insider threats [43]. Based on the aforementioned, we hypothesise:

*H 4:* Increasing the risk for information security misconduct positively influences employees' attitude towards preventing delinquent behaviour in the domain of information security.

## 3.4. Reduce the Rewards

Rewards are the extrinsic motivation for individuals' behaviour in many cases. Rewards encourage them to engage in a particular behaviour [44]. 'Reducing the rewards' refers to the benefit of the crime in this research, particularly when employees sell organisational information assets. Beebe and Rao [45] showed that sanctions are not enough to discourage offenders from committing crimes, and that the benefits of their violations should be reduced as an effective approach to dissuade them from conducting crime; the perception of minimal benefit by offenders discourages them from perpetrating crime. Encryption (data deformation), watermarking (identifying property), information and hardware segregation (removing target), and minimising reconnaissance information (concealing targets), are examples of methods that reduce the benefits for employees who engage in information security misbehaviour [33]. A digital signature, which shows the validity and integrity of a document that can be used, as well as other methods, such as time stamps, reduces the benefit for offenders [46]. Li, Zhang [29] offered automatic data destruction mechanisms and insider continuity management as an effective approach that mitigates benefits for offenders. Based on the above we conjecture that:

*H 5:* Reducing the rewards for information security misconduct positively influences employees' attitude towards preventing delinquent behaviour in the domain of information security.

### 3.5. Reduce Provocations

Provocation refers to the action or occurrence that causes someone to do something or become angry. Provocation is a stimuli for individuals' behaviour, they show negative and aggressive behaviour under such conditions [47]. By reducing provocation, we try to reduce the emotional causes and motivation for conducting an offence. Managing negative issues and preventing disputes in the working environment, decreasing emotional arousal, frustration and stress, discouraging imitation and neutralising peer pressure are examples of provocation reduction techniques in organisations [23, 48]. Silowash, Cappelli [49] asserted that controls and security policies can be misunderstood due to poor communication or inconsistently applied; employees' involvement in the process of development and implementation of information security is a useful approach to counter this issue [50]. Security usability could also influence the insider's negative response towards information security control. Anger, fear, guilt, happiness and joy are other factors that affect employees' attitude towards misbehaviour in the domain of information security; management should reduce any provocations that threaten information security in organisations [51]. Hence, the following hypothesis is proposed in this research:

*H 6:* Reducing provocations for information security misconduct positively influences employees' attitude towards preventing delinquent behaviour in the domain of information security.

### 3.6. Remove Excuses

Rationalisation and justification of misconduct plays an important role in the formation of crime. Rationalisation or making excuses is a defence mechanism to justify and explain a violation in a logical and rational manner. Miscreants even try to present their misconduct as being tolerable or admirable by rationalisation [52]. It is acknowledged that rationalisation influences the violation of organisational information security policies [27]. Sharing a network password for convenience, and justification thereof by contending that nobody will be injured by this action, is an example of wrong rationalisation. This kind of rationalisation has a negative impact on employees' behaviour, and even causes employees to knowingly deviate from security policies. They endeavour to decrease their shame and guilt at deliberately violating IT policies by rationalising their motivations. They try to present their misconduct as being more normal and necessary than it actually is [39]. Providing clear documents, controlling and monitoring, and consistently enforcing policies are approaches that can inhibit the practice of making such excuses by individuals. Clarification of information security rules and policies, cyber ethics training, alerting conscience and assisting employees in complying with OISPs are other examples of this approach to removing excuses from staff. Hence, the following hypothesis is proposed:

*H 7:* Removing excuses for information security misconduct positively influences employees' attitude towards preventing delinquent behaviour in the domain of information security.

### 3.7. Attitude, Perceived Behavioural Control and Subjective Norms

The past and present experience of individuals influences their attitude, where by 'attitude' we intend to refer to the favour or disfavour towards a subject such as an idea, event, a person or other object [8]. In simple terms, attitude is the result of an individual's evaluation concerning a subject in question, ranging from extremely bad to extremely good. Attitude also relates to people's negative or positive views towards conducting a specific behaviour. Hepler [53] believed that attitude is a psychological status that is formed based on the individual's stimuli. Attitude influences an individual's behaviour. The set of beliefs that a person has about an object affects his or her attitude, intention, and, ultimately, his or her behaviour. Siponen, Adam Mahmood [11] showed that an employee's attitude influences their behaviour to comply with OISPs. Jeon, Kim [54] revealed that a positive attitude about knowledge sharing significantly influences individuals' behaviour towards sharing their knowledge. Therefore, we postulate that:

*H 8:* Attitude influences employees' intention towards preventing delinquent behaviour in the domain of information security.

The perceived social pressure to perform or not perform a behaviour manifests subjective norms [55]. Subjective norms are the effect of individuals' opinions about a particular behaviour [22]. Protection of information assets is important to management, heads of department, supervisors, colleagues, or, in other words, significant others. Subjective norms affect employees' intentions towards preventing information security misconduct [23]. In this research we postulate that:

*H 9:* Subjective norms influence employees' intention towards preventing delinquent behaviour in the domain of information security.

One of the important factors in the TPB is perceived behavioural control, by which we mean the perception of the hardness or easiness of performing a behaviour or task on the part of the individual [55]. Safa, Sookhak [16] showed that perceived behavioural control influences the formation of information security conscious care behaviour. In this research, perceived behavioural control relates to the belief that engaging in information security behaviour and preventing information security misconduct are not difficult tasks. All employees are able to engage in proper information security behaviour. This is why we present the hypothesis below:

*H 10:* Perceived behavioural control influences employees' intention towards preventing delinquent behaviour in the domain of information security.

### 3.8. Intention

Intention is an important element in the formation of behaviour. Intention represents a commitment to carry out an action, either now or in the future. Intention contains the concept of planning and forethought [56]. Based on the TPB, attitude, perceived behavioural control and subjective norms play key roles in the creation of intention in order to achieve goals [34]. In other words, a desire towards achieving a goal that satisfies a person's generates an intention to engage in behaviour that promotes that goal in him or her. Shropshire, Warkentin [8] revealed that intention significantly affects the adoption of information security behaviour in organisations. Hence, the following hypothesis is proposed:

*H 11:* Intention to prevent misbehaviour mitigates insider threats in organisations.

Figure 2 depicts the conceptual model and hypotheses in a concise form.

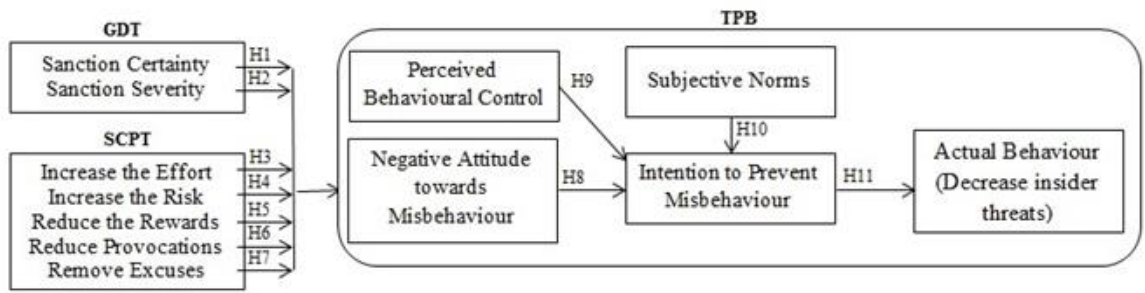

**Figure 2: Conceptual framework**

Table 1 shows definition of the factors in the conceptual framework.

Table 1: Definition of factors in the research model

| Theories | Constructs | Definitions in this research |
|---|---|---|
| **General Deterrence Theory** (employees' perception) | Sanction Certainty | Refers to the belief that the authority will detect his or her delinquent behaviour. |
| | Sanction Severity | Refers to the belief that the authority will consider a punishment, such as fine, dismissal or even jail based on the effect of his or her delinquent behaviour. |
| **Situational Crime Prevention Theory** (environmental factors-opportunity reduction) | Increase the Effort | Refers to difficulty of committing a delinquent behaviour, which may dissuade offender from conducting crime. |
| | Increase the Risk | Refers to the consequence of delinquent behaviour, such as job termination. |
| | Reduce the Rewards | Refers to the decreasing benefits or revenue of the delinquent behaviour. |
| | Reduce Provocations | Refers to mitigating or removing noxious stimuli, such as conflict, unnecessary stress or competition from the workplace. |
| | Remove Excuses | Refers to removing the rationalisations of the delinquent behaviour. |
| **Theory of Planned Behaviour** (behaviour formation) | Attitude | Refers to an expression of disfavour or favour towards an object, such as secure information behaviour. |
| | Perceived Behavioural Control | Refers to the difficulty of the behaviour (secure information behaviour). |
| | Subjective Norms | Refers to performing or not performing the behaviour. |
| | Intention | Represents a commitment to act with forethought and planning now or in future. |
| | Actual Behaviour | Refers to the mitigation of insecure information behaviour (insider threats) in organisations. |

## *4. Research methodology*

This study strives to show how management can mitigate the risk of insider threats by focusing on a preventative approach in organisations. A literature review from high quality journals, besides the theoretical background and expert views increases the reliability of the conceptual model. The research model was improved by expert feedback and by use of the Delphi method. The framework was improved using qualitative and quantitative methods. The data was collected from several organisations in the UK. A questionnaire using Likert scales was used for data gathering.

The model was created based on the literature review and theoretical background. That is why Confirmatory Factor Analysis (CFA) was considered in order to determine whether the measurement model (MM) confirms our understanding of the constructs. In simple terms, whether our hypotheses are confirmed by the data that we have collected. Structural Equation Modelling (SEM) has been acknowledged as a suitable method to investigate the relationships among the independent, mediating and dependent variables in such a model [57]. The Maximum Likelihood method in IBM Amos 20 was used to assess the measurement and structural models [58]. The other statistical measurements that demonstrate the reliability of the conceptual model have presented in Table 5.

*4.1.* **Data collection**

Data gathering was conducted on the employees of several companies that are active in the domain of e-Commerce, banking and education. The questions in the questionnaire were developed on the basis of the framework structure and the concepts of factors. In this step, we also considered previous similar studies and adopted questions therefrom. To reply to the questions, a range of options from strongly agree to strongly disagree (Likert Scales) was used. We explained the aims of this study to participants and kindly requested that they answer the questions on the bases of their experience and opinion. The consent of respondents to participate in this research was important to us; after indication of their consent, we asked them to start answering the questions. We confirmed that this data would be only used for academic purposes and kept confidential.

Whether the questions were applicable, comprehendible and subject to a single interpretation on the part of the respondents would have a significant effect on the results. That is why we pilot-tested the questionnaire with 42 participants before distribution. We looked at their hesitation, emotions and descriptions during the pilot test. Based on their reaction and comments, we revised and improved some of the questions to increase the reliability of the questionnaire. The last version of the questionnaire included 51 questions, each factor was indicated by various items (questions). Table 3 shows in a clear manner how every factor is measured using several items.

*4.2.* **Demography**

Data collection is usually a time-consuming process; we used two approaches to data collection – a paper-based questionnaire and an electronic questionnaire – in order to expedite the procedure. The questionnaire was hosted on Google Drive and was emailed to employees for whom we had email addresses. Four hundred and eighty-six respondents answered the questions, of which 152 used the paper-based questionnaire and 334 used Google Drive. We immediately reviewed the responses and asked the respondents to kindly complete the questions to which they have not replied, thereby decreasing the number of incomplete questions in the paper-based questionnaire. Nevertheless, nine questionnaires (5.2%) were discarded due to incomplete answers, or because the respondent replied to all the questions in a similar manner.

Google Form helped us to distribute the questionnaire effectively and efficiently through the Internet. The electronic questionnaire was emailed to those employees for whom we had email addresses. Thirty-three electronic questionnaires were discarded from three hundred and thirty-four, due to incomplete responses or inappropriate status. Finally, four hundred and forty-four responses were considered and transferred to the main dataset for data analysis. Table 2 shows the demography of the participants.

| Table 2: Participants' characteristics | | | |
|---|---|---|---|
| **Measure** | **Items** | **Frequency** | **Per cent** |
| *Gender* | Male | 246 | 55.4 |
| | Female | 198 | 44.6 |
| *Age* | 21 to 30 | 116 | 26.1 |
| | 31 to 40 | 198 | 44.61 |
| | 41 to 50 | 81 | 18.25 |
| | Above 50 | 49 | 11.04 |
| *Position* | Employee | 398 | 89.64 |
| | Chief employee | 36 | 8.11 |
| | Management | 10 | 2.25 |
| *Work experience* | 1 to 2 years | 98 | 22.1 |
| | 3 to 5 years | 222 | 50 |
| | Above 5 years | 124 | 27.9 |
| *Education* | Diploma | 36 | 8.1 |
| | Bachelor | 298 | 67.12 |
| | Master | 101 | 22.75 |
| | PhD | 9 | 2.03 |

## 5. Results

The research variables are usually unquantifiable and unobservable (latent), and are usually measured by several items, such as perceived sanction certainty and severity, effort, risk and so forth. The MM and SM are two important parts of data analysis in Structural Equation Modelling (SEM) that can be used to show the validity and reliability of the research model. The MM displays the relationship between the variables (items) and the main factors. In other words, the MM shows that these items measure the relevant factor appropriately. The reliability and validity of the observed variables (items) were tested before the MM was fitted to the data. The SM investigates the relationship between the unobservable variables (factors). SEM is the most appropriate method for this kind of research model [58].

### 5.1. Measurement model

SEM explores the relationship among the variables and confirms or rejects the hypotheses. SEM not only estimates the regression among the latent variables, but also isolates the error when it measures the latent variables. The normality of data distribution shows what kinds of tests should be used in data analysis; that is why skewness and kurtosis tests were used in the first step of data analysis. The results were between -2 and +2, which shows a normal distribution [59]. The research model was developed based on the literature review with a theoretical background, which is why confirmatory factor analysis (CFA) was considered to be a suitable approach for this research. CFA investigates whether the measured variables are consistent with our understanding of the variables and factors in the research model [60].

Convergent validity was explored using factor loading of the variables (items). A factor loading of more than 0.5 shows acceptable convergent validity [58]. The items with a factor loading of less than 0.5 were discarded from the research model. The IR3 in the Increase the Risk, RP2 in the Reduce Provocation, RE4 from Remove Excuses and PBC3 from Perceived Behavioural Control were extracted from the model due to their lesser factor loading on the related constructs. Cronbach's Alpha indicates the internal consistency and shows the correlation among the items (observable variables) used to measure a factor (unobservable variables). A Cronbach's Alpha with a measure more than 0.7 indicates acceptable internal consistency for the model [61]. Some of the statistical measures that relate to factors and the items that measure them have been presented in Table 3.

**Table 3: The factors, items, and their descriptive statistics**

| Construct | | Items | Mean | Std Dev | CFA Loading | Composite reliability |
|---|---|---|---|---|---|---|
| **Perceived Sanction Certainty (PSC)** | PSC1 | I believe that if I violate confidentiality of information the management will realise it. | 3.92 | .78 | .612 | .816 |
| | PSC2 | I believe that if I transfer organisational information outside the management will find out my violation. | 4.01 | .82 | .714 | |
| | PSC3 | I believe that if I sell organisational information my organisation will discover it. | 4.12 | .76 | .592 | |
| | PSC4 | I believe that if I do not comply with OISPs and procedures my boss will detect it. | 4.08 | .92 | .696 | |
| **Perceived Sanction Severity (PSS)** | PSS1 | I think the consequences of the violation of OISPs are very bad for me. | 4.06 | 1.01 | .648 | .786 |
| | PSS2 | I deserve punishment if I violate the confidentiality of organisational information. | 3.82 | .92 | .724 | |
| | PSS3 | I think punishment will be high if I sell or transfer organisational information outside. | 4.16 | .82 | .764 | |
| | PSS4 | I think receiving sanctions because of my information security misconduct will negatively influence my career development. | 3.96 | .76 | .623 | |
| **Increase the Effort (IE)** | IE1 | Control of information access affects my attitude to be careful about my information security behaviour. | 3.86 | .88 | .722 | .698 |
| | IE2 | Trying to pass authentication systems influences my attitude to prevent misbehaviour. | 4.02 | .92 | .762 | |
| | IE3 | Access to isolated sensitive information needs more effort that influences my attitude to prevent misconduct. | 4.12 | .82 | .742 | |
| | IE4 | Surveillance on employees' access to information affects my attitude to prevent violation of information policies. | 3.98 | .84 | .816 | |
| **Increase the Risk (IR)** | IR1 | Tracking my access to information on the systems affects my attitude to prevent information security misconduct. | 4.21 | .92 | .722 | .716 |
| | IR2 | Reducing anonymity influences my attitude to avoid information security misbehaviour. | 3.98 | .79 | .736 | |
| | IR3 | Monitoring and controlling access to information influences my attitude to be careful about my behaviour. | 4.28 | .76 | Dropped | |
| | IR4 | Possibility of identification by management influences my attitude to avoid information security misconduct. | 4.04 | .84 | .698 | |
| **Reduce the Rewards (RR)** | RR1 | Automatic data destruction eliminates benefits of information for offenders and dissuades them from misbehaviour. | 4.02 | .92 | .668 | .792 |
| | RR2 | Encryption of data removes benefits of information and prevents information security misconduct. | 4.11 | 8.86 | .748 | |
| | RR3 | Watermarking eliminates personal benefits and prevents information security misbehaviour. | 3.94 | 1.02 | .764 | |

| Construct | Item | Statement | Mean | SD | Loading | α |
|---|---|---|---|---|---|---|
| | RR4 | Elimination of benefits influences employees' attitude to prevent information security misconduct in organisations. | 4.04 | .89 | .769 | |
| **Reduce Provocations (RP)** | RP1 | Avoiding disputes reduces provocation and positively influences my attitude to avoid misbehaviour. | 4.12 | .78 | .746 | .806 |
| | RP2 | Reducing my stress decreases provocation for information security misbehaviour. | 3.98 | .86 | Dropped | |
| | RP3 | Elimination of employees' frustration mitigates provocation for information security misbehaviour. | 4.16 | .82 | .724 | |
| | RP4 | Reducing emotional arousal decreases provocation and positively influences my attitude to avoid misconduct. | 3.86 | .78 | .782 | |
| | RP5 | I believe reducing provocations in organisations positively influences my attitude to avoid misbehaviour. | 4.13 | .91 | .728 | |
| **Remove Excuses (RE)** | RE1 | Clarification of information security policies positively influences my attitude to avoid misbehaviour. | 4.02 | 1.04 | .746 | .726 |
| | RE2 | Cyber ethics training positively influences my attitude to avoid misbehaviour. | 3.96 | .86 | .821 | |
| | RE3 | Assisting compliance with organisational information security policies positively influences my attitude to avoid misbehaviour. | 4.16 | .92 | .764 | |
| | RE4 | Alerting employees' conscience positively influences my attitude to avoid misbehaviour. | 4.04 | .83 | Dropped | |
| | RE5 | Removing excuses from organisational environment positively affects my attitude to avoid misbehaviour. | 3.98 | .86 | .804 | |
| **Attitude (AT)** | AT1 | Safe information security behaviour protects information assets in organisations. | 4.04 | .81 | .726 | .684 |
| | AT2 | Appropriate information security behaviour mitigates the risk of information security breaches in organisations. | 4.16 | .92 | .748 | |
| | AT3 | Safe information security behaviour decreases information security incidents in organisations. | 4.06 | .84 | .728 | |
| | AT4 | Proper information security behaviour is a good practice. | 4.18 | .78 | .722 | |
| **Perceived Behavioural Control (PBC)** | PBC1 | I have the necessary abilities to have safe information security behaviour. | 3.94 | .92 | .768 | .748 |
| | PBC2 | I am able to mitigate information security threats in my organisation. | 4.14 | .84 | .726 | |
| | PBC3 | Safe information security behaviour is an easy task for me. | 3.94 | .89 | Dropped | |
| | PBC4 | I have enough knowledge to behave safe in terms of information security. | 4.12 | 1.01 | .546 | |
| **Subjective Norms (SN)** | SN1 | My colleagues think that we should behave safe to protect organisational information assets. | 4.18 | .92 | .688 | .802 |
| | SN2 | The head of department believes that we should protect organisational information assets. | 3.82 | .94 | .592 | |
| | SN3 | The senior staff in my company have a positive view about the protection of information by employees. | 4.01 | 1.03 | .728 | |
| | SN4 | My friends in my office encourage me to have safe information security behaviour. | 4.12 | .82 | .684 | |
| **Intention (IN)** | IN1 | I am willing to safeguard organisational information assets. | 3.86 | .96 | .628 | .782 |

| | | | | | |
|---|---|---|---|---|---|
| | IN2 | I intentionally help my colleagues to increase information security. | 4.08 | .92 | .728 |
| | IN3 | I collaborate with other staff to decrease insider threats in my organisation. | 4.12 | .85 | .698 |
| | IN4 | I plan to have safe information security behaviour. | 4.04 | .92 | .592 |
| *Actual Behaviour (AB)* | AB1 | I try to avoid mistakes in the domain of information security. | 3.92 | .86 | .738 |
| | AB2 | I always try to mitigate information security threats. | 4.08 | 1.02 | .766 |
| | AB3 | I think about the consequences of my behaviour before any action. | 3.89 | .96 | .686 | .812 |
| | AB4 | I am careful about my behaviour in the domain of information security. | 4.14 | .88 | .594 |
| | AB5 | I frequently asses my information security behaviour to improve it. | | | |

OISPs: Organisational Information Security Policies
Factor loading from confirmatory factor analysis.
t-value is significant at p < 0.05

Different factors were linked to another in order to be assured about convergent and discriminant validity of the model. The factors are independent and unique. Convergent validity shows whether there is any relationship between factors in the model and with each other. Discriminant validity investigates the lack of correlation between factors that they should not have relationship in the model. Table 4 shows the correlation between different constructs [58].

**Table 4: Correlation between different constructs**

| | | Mean | SD | 1 | 2 | 3 | 4 | 5 | 6 | 7 | 8 | 9 | 10 | 11 | 12 |
|---|---|---|---|---|---|---|---|---|---|---|---|---|---|---|---|
| 1 | **PSC** | 4.04 | 0.94 | 0.826 | | | | | | | | | | | |
| 2 | **PSS** | 4.12 | 0.82 | 0.402 | 0.848 | | | | | | | | | | |
| 3 | **IE** | 4.08 | 0.78 | 0.304 | 0.422 | 0.779 | | | | | | | | | |
| 4 | **IR** | 4.18 | 1.02 | 0.468 | 0.346 | 0.422 | 0.798 | | | | | | | | |
| 5 | **RR** | 4.06 | 1.04 | 0.487 | 0.424 | 0.437 | 0.265 | 0.896 | | | | | | | |
| 6 | **RP** | 4.12 | 0.96 | 0.498 | 0.252 | 0.258 | 0.286 | 0.221 | 0.887 | | | | | | |
| 7 | **RE** | 4.14 | 0.98 | 0.248 | 0.514 | 0.362 | 0.266 | 0.432 | 0.494 | 0.822 | | | | | |
| 8 | **AT** | 4.22 | 1.02 | 0.612 | 0.522 | 0.521 | 0.716 | 0.695 | 0.546 | 0.536 | 0.868 | | | | |
| 9 | **PBC** | 4.26 | 1.14 | 0.188 | 0.234 | 0.198 | 0.247 | 0.226 | 0.288 | 0.368 | 0.442 | 0.724 | | | |
| 10 | **SN** | 4.04 | 0.86 | 0.438 | 0.538 | 0.623 | 0.636 | 0.248 | 0.506 | 0.484 | 0.368 | 0.564 | 0.829 | | |
| 11 | **IN** | 4.14 | 1.18 | 0.356 | 0.366 | 0.253 | 0.184 | 0.198 | 0.282 | 0.268 | 0.623 | 0.639 | 0.562 | 0.836 | |
| 12 | **AB** | 4.02 | 0.98 | 0.204 | 0.218 | 0.224 | 0.329 | 0.248 | 0.198 | 0.326 | 0.348 | 0.336 | 0.394 | 0.644 | 0.746 |

## 5.2. Testing the structural model

Structural Equation Modelling (SEM) applies different statistical tests to examine a theoretical model or conceptual framework. SEM not only investigates all relationships between different variables, but also isolates observational errors from the measurements of latent variables. SEM tests the overall data fit to the model and presents reliable measurement. IBM AMOS version 20 is the statistical software that has been used in this research.

A review of literature helped us to develop the research model and the entire model has been covered by three basic theories, so that the reliability of the model is increased. For this reason, Confirmatory Factor Analysis (CFA) was applied instead of Exploratory Factor Analysis (EFA). Fit indices play important roles regarding the validity of the model; Comparative and Global fit measures were applied to investigate fit indices. Table 5 displays the model fit indices in a concise format.

**Table 5: Model fit indices**

| Fit indices | Model value | Acceptable standard |
|---|---|---|
| $\chi^2$ | 1002.62 | - |
| $\chi^2/Df$ | 1.92 | <2 |
| GFI | 0.926 | >0.9 |
| AGFI | 0.964 | >0.9 |
| CFI | 0.933 | >0.9 |
| IFI | 0.908 | >0.9 |
| NFI | 0.942 | >0.9 |
| RMSEA | 0.076 | <0.08 |

The results of the data analysis are presented in Table 6. The findings showed that the paths from perceived sanction certainty (β=0.722, p=0.005), perceived sanction severity (β=0.789, p=0.004), increase the effort (β=0.642, p=0.011), increase the risk (β=0.522, p=0.021), reduce the rewards (β=0.703, p=0.007) towards safe information security attitudes were significant. However, the effect of reducing the provocation and removing excuses towards attitudes were not significant. Therefore, H6 and H7 are rejected. The findings also revealed that attitude (β=0.685, p=0.009), perceived behavioural control (β=0.561, p=0.019), and subjective norms (β=0.726, p=0.001) towards intention to secure information behaviour were significant. Finally, the results showed that the intention to protect information security behaviour (β=0.798, p=0.001) had significant effects on the employees' behaviour towards mitigating insider threats in organisations.

**Table 6: The results of the hypotheses testing**

| Path | | | Standardized estimate | p-Value | Results |
|---|---|---|---|---|---|
| PSC | ⟶ | AT | 0.722 | 0.005 | Support |
| PSS | ⟶ | AT | 0.789 | 0.004 | Support |
| IE | ⟶ | AT | 0.642 | 0.011 | Support |
| IR | ⟶ | AT | 0.522 | 0.021 | Support |
| RR | ⟶ | AT | 0.703 | 0.007 | Support |
| RP | ⟶ | AT | 0.598 | 0.064 | Not-Supported |
| RE | ⟶ | AT | 0.424 | 0.056 | Not-Supported |
| AT | ⟶ | IN | 0.685 | 0.009 | Support |
| PBC | ⟶ | IN | 0.561 | 0.019 | Support |
| SN | ⟶ | IN | 0.726 | 0.001 | Support |
| IN | ⟶ | AB | 0.798 | 0.001 | Support |

## 6. Contribution and implementation

The significant aspect of this study is derived from the inclusion of the deterrence and crime prevention approaches that are the results of two basic theories – Deterrence and Situational Crime Prevention Theory. The presented factors dissuade employees from information security misconduct in organisations, and, consequently, mitigate insider threats. Both theories have the same effect on individuals' attitudes, but the GDT emphasises the individual's perception and attitude, and Situational Crime Prevention Theory highlights the environmental restrictions which function to mitigate insider threats.

To the best of our knowledge, this is among the first studies to conceptualise insider threat prevention on the bases of prevention and deterrence. This synthesis constitutes a new perspective which enables organisations to better manage insider threats. We believe that this complements the previous studies that have been carried out in this domain.

The output of statistical analysis revealed that perceived sanction certainty and severity influence individuals' attitudes towards preventing information security misconduct in organisations. This finding is in-line with the output of Cheng, Li [23]. The results also showed that increasing the effort, risk and reducing the rewards significantly influences employees' attitudes towards preventing information security misbehaviour. A plausible reason for this finding might be the high risk and low benefit of their misconduct that affects their final decision to prevent information security misbehaviour. Contrary to our expectation, reducing provocation and excuses did not significantly affect an individual's attitude towards preventing information security misconduct. One conceivable explanation for this finding might be in the culture of the people in the UK. Moral values are important in their culture, and personal affairs do not influence their duties in the work place. The results also showed that a negative attitude towards information security misbehaviour, perceived behavioural control (belief that having safe information security behaviour is an easy task), and personal norms (belief that information security misconduct is a negative behaviour), all influence employees' intention to engage in information security misbehaviour. Indeed, these factors originate from the Theory of Planned Behaviour that has been applied in many studies previously in this domain [4, 16, 22]. The results of the statistical analysis and the review of the literature demonstrate the soundness and effectiveness of the proposed model.

## 7. Conclusion, Limitations and future work

Information technology has changed organisational activities so as to make them become faster, and more effective and efficient. However, protection of information is still a challenging subject for all companies. Anecdotal and empirical evidence has shown that insider threats are responsible for a significant portion of the risk in the domain of information security [43, 62]. This research endeavours to improve and diversify research on information security insider threats in organisations through the Deterrence and Situational Crime Prevention

Theories. Factors, such as perceived sanction certainty and severity, increasing the effort and risk for information security misconduct, and reducing the rewards have a significant effect on employees' attitude towards preventing misbehaviour. In addition, a negative attitude towards information security misconduct, perceived behavioural control and personal norms influence individuals' intention, and, ultimately, their behaviour in order to mitigate insider threats in organisations.

The research model encompasses three main sections. The first part relates to the employees' perception of sanctions. The second part refers to the restrictions and controls (environmental factors), such as increasing the effort and risk, decreasing the rewards and provocations, and removing excuses. Finally, the third part shows how mitigation of insider threats forms in organisations. Looking at the model, it can be seen that insider threat is a managerial issue and controllable. It is clear that insider threats can be managed through psychological, managerial and technological aspects regarding information security.

To extend this research, we can look at the problem statement (insider threats) from different perspectives; this research can be continued further by focusing on the role of organisational values and culture. Moral values dissuade individuals from misconduct. Another clue for developing this research is the effect of organisational bonds, such as attachment to one's organisation, involvement in information security, commitment to organisational policies and plans, and, finally, personal norms such as the norm that having safe information security behaviour is a positive factor and the norm that information security misconduct a negative behaviour. Motivation for crime is an important factor in delinquent behaviour. Intrinsic and extrinsic motivation can also be the focus of future research in this domain.

This research faced several limitations. OISPs play an important role in the mitigation of information security breaches. We tried to collect data from organisations that had established OISPs, as employees in such companies are aware of the importance of information security. They can better understand the purpose of this study and the concepts that are used in the questionnaire. Unfortunately, there is a paucity of such companies in the UK. Collecting data in the domain of information security, even in non-military organisations, is a difficult task. The data was collected from companies from which we obtained permission for data collection. The precision and generalisation of the results can be improved with a bigger sample size and by increasing the number of companies investigated. If possible, data collection can also be extended to other countries in future research. The data was gathered by Google Drive which is sensitive, as it is based on participants' email addresses. This means that participants with more than one email address can answer the questionnaire two or more times. Although the probability of participation more than once is almost zero, we would have operated with a facility to check this problem or check their IP address to detect them. In this way we would have been able to recognise participants with two or more responses.